\documentclass[final,5p,times,twocolumn]{elsarticle}

\usepackage{amssymb}
\usepackage{marvosym}
\usepackage{mnsymbol}
\usepackage{pifont}
\usepackage{bm}

\usepackage[nodots,nocompress]{numcompress}

\def\etal{{\it et al.}}

\journal{Nuclear Instruments and Methods in Physics Research B}

\begin{document}

\begin{frontmatter}

\title{PEBSI - A Monte Carlo simulator for bremsstrahlung arising from electrons colliding with thin solid-state targets}

\author[aff1,aff2]{G.~Weber}
\ead{g.weber@gsi.de}
\author[aff2,aff3]{R.~M{\"a}rtin}
\author[aff2,aff3]{A.~Surzhykov}
\author[aff4]{M.~Yasuda}
\author[aff5,aff2,aff3]{V.~A.~Yerokhin}
\author[aff1,aff2,aff3]{Th.~St\"ohlker}
\address[aff1]{Helmholtz-Institut Jena, Jena, Germany}
\address[aff2]{GSI Helmholtzzentrum f{\"u}r Schwerionenforschung GmbH, Darmstadt, Germany}
\address[aff3]{Physikalisches Institut, Ruprecht-Karls-Universit{\"a}t, Heidelberg, Germany}
\address[aff4]{Department of Physics and Electronics, Osaka Prefecture University, Osaka, Japan}
\address[aff5]{St. Petersburg State Polytechnical University, St. Petersburg, Russia}

\address{}

\begin{abstract}
We present a Monte Carlo code dedicated to the simulation of bremsstrahlung arising in collisions of polarized electrons with thin target foils. The program consists of an electron transport algorithm taking into account elastic electron-nucleus scattering and inelastic collisions with target electrons as well as a treatment of polarized-electron bremsstrahlung emission. Good agreement is found between the predictions of the electron transport code and data stemming from other simulation programs and experiments. In addition, we present first results from the bremsstrahlung simulation which indicate a significant decrease in the degree of linear polarization of bremsstrahlung even for the thinnest gold targets considered.
\end{abstract}


\end{frontmatter}


\section{Introduction}
Bremsstrahlung arising from the interaction of an energetic electron with a screened nucleus potential, also referred to as ordinary bremsstrahlung, is one of the basic photon-matter processes and has attracted continuous interest both theoretically and experimentally during the last decades~\cite{nak94,qua00,qua06}. Of particular importance is the study of the bremsstrahlung process resulting from polarized electrons as it reveals subtle details of the polarization transfer from charged particles to photons. The dependency of bremsstrahlung on the electron spin can be expressed in terms of the so-called polarization correlations that have been systematically studied by Tseng and Pratt~\cite{tseng73} and were recently revisited in several theoretical works~\cite{yer10,jak11a,jak11b}. Recent interest focussed on the polarization properties of the emitted bremsstrahlung photons with respect to the incoming electron polarization. In general, the use of polarized electrons should lead to a significant change in the degree of linear polarization and a rotation of the polarization axis of the bremsstrahlung photons with respect to the unpolarized electron case, where the photon polarization is solely determined by the kinematic and atomic parameters of the interaction. It is interesting to note, that a similar effect was recently predicted for the radiative recombination process in case of polarized highly-charged, heavy ions, see~\cite{sur05} for details. These theoretical studies were mainly motivated by the development of novel Compton polarimeters that enable efficient and precise measurements of linear polarization in the hard x-ray regime~\cite{tas06,spill08,web10a,web10b}.

During the last two years, such polarimeters were applied in a series of polarization-resolved bremsstrahlung measurements that have been carried out at the teststand of the polarized electron source SPIN~\cite{hes08,yul11} at the Technical University of Darmstadt. In these experiments the polarization transfer was studied in collisions of polarized electrons with gold foils of about 100\,$\mu$g/cm$^2$ thickness and at an impact energy of 100\,keV~\cite{mar11,stan11}. This kind of investigations may open a route for 'complete' measurements where besides the energy and angular information also the polarization properties of the involved particles are obtained~\cite{nak94}. However, a crucial point when drawing conclusions from the experimental data is the question to what extend the bremsstrahlung characteristics, in particular concerning the polarization, is altered by effects due to multiple collisions inside the target foil. While the influence of the target thickness on bremsstrahlung angular and spectral distribution was intensively investigated in several works~\cite{pla67,wil06,shan06,har07,wil08a,wil08b}, to the best of our knowledge no systematic study of target effects on the bremsstrahlung polarization was ever published. However, one can expect that the linear polarization shows a high sensitivity to target effects as the straggling of incident electrons followed by bremsstrahlung emission will lead to a superposition and, consequently, to a partial cancelation of different polarization orientations in the detector.

In general, Monte Carlo simulations are well-suited for the determination of multiple-scattering effects on the properties of processes where rigorous theoretical treatments assume single-collision conditions. However, in case of bremsstrahlung most of the general-purpose Monte Carlo programs available (\mbox{PENELOPE}, \mbox{Geant4}, \mbox{EGS5}, etc.) use approximations that are limited to thick-target bremsstrahlung and/or ignore the effects of electron polarization on the bremsstrahlung properties.

In this work we present the Monte Carlo code PEBSI (\underline{P}olarized \underline{E}lectron \underline{B}remsstrahlung \underline{SI}mulator) that models the transport of polarized electrons through the target material in order to estimate the bremsstrahlung properties resulting from the finite thickness of the target foils. Here, the bremsstrahlung theoretical data are taken from fully relativistic calculations taking into account electron and photon polarization. The development of the code was originally motivated by bremsstrahlung polarimetry measurements as described above. However, it might be applicable to a variety of studies dealing with polarized electron transport and bremsstrahlung emission.

\section{Polarized electron transport model}
In the energy region between a few keV up to a few MeV the electron transport in solid state targets is governed by two processes, namely the elastic electron-nucleus scattering and the inelastic scattering on bound target electrons, with the latter leading to a successive energy loss of the incident electrons. In the case of high-Z targets, the $Z^2$ scaling of the scattering cross sections leads to a dominance of the elastic scattering process. Thus, for our application the elastic scattering cross section should be treated as accurately as possible, while for the inelastic scattering even a rather approximate handling will not significantly decrease the validity of the electron transport model~\cite{pan89,mur95,ivi03}. Note that we do not consider the bremsstrahlung process within the electron transport code as its contribution is negligible in the electron energy region of interest. Consequently, the mean free path $\lambda$ between two interactions in the target is given by

\begin{equation}
\frac{1}{\lambda}=\rho\left(\sigma_{\text{el}}+\sigma_{\text{inel}}\right)\,,
\label{eq1}
\end{equation}

where $\rho$ is the target density and $\sigma_{\text{el}}$ and $\sigma_{\text{inel}}$ denote the elastic and inelastic interaction cross sections per atom, respectively. The incident electrons are followed as long as they do not leave the target foil (transmission or backscattering) and their kinetic energy stays above a certain cut-off value $E_\text{cut}$.

The general structure of our Monte Carlo program is adopted from the approach presented in~\cite{shi92}. In the following we briefly discuss the implementation of both processes from Eq.~\ref{eq1} in the PEBSI code. In addition, results of the electron transport code are compared to data from experiments and a different Monte Carlo program.

\subsection{Elastic electron-nucleus scattering}
Here, we summarize the main aspects of the elastic electron-nucleus scattering process as it is implemented in the PEBSI code. For a much more detailed discussion we refer the reader to~\cite{kes85}.

In case of unpolarized electrons the differential cross section is given by the Mott equation~\cite{mott29}:

\begin{equation}
{\left( \frac{\text{d}\sigma_{\text{el}}}{\text{d}\Omega} \right)}_{0} =  {|f\!\left(\theta\right)|}^2+{|g\!\left(\theta\right)|}^2\,,
\label{mott1}
\end{equation}

where $\theta$ denotes the polar scattering angle and $f$ is the amplitude of the scattering wave with spin direction remaining unchanged while $g$ is the amplitude of the spin-flip scattering wave as a result of the spin-orbit coupling.

When polarized electrons are considered an additional dependence on the azimuthal scattering angle~$\phi$ is introduced:

\begin{equation}
\frac{\text{d}\sigma_{\text{el}}}{\text{d}\Omega} =  {\left( \frac{\text{d}\sigma_{\text{el}}}{\text{d}\Omega} \right)}_{0} \left(1+ S\!\left(\theta\right) \bm{P} \cdot \bm{\hat n}\right)\,,
\label{mott2}
\end{equation}

where $\bm{P}$ denotes the orientation of the incident electron spin and $\bm{\hat n}$ is the unit vector perpendicular to the scattering plane, which is defined by the directions of the incoming and the outgoing electron. $S\!\left(\theta\right)$ is called the Sherman function and is given by

\begin{equation}
S\!\left(\theta\right)=i\frac{f\,g^{*}-f^{*}\,g}{{|f\!\left(\theta\right)|}^2+{|g\!\left(\theta\right)|}^2}
\label{sherm}
\end{equation}

with $^{*}$ indicating the complex conjugate. According to Eq.~\ref{mott2}, incident electrons with a spin orientation having a transversal component with respect to the electron momentum will exhibit an anisotropic azimuthal scatter distribution with the degree of this asymmetry being determined by the value of $S\!\left(\theta\right)$. This effect is used in Mott polarimetry where the degree of electron spin polarization is obtained by measuring the elastically scattered electron intensity distribution, see~\cite{gay92} for details.

After the scattering process took place the new orientation of the electron spin $\bm{P^\prime}$ is given by

\begin{equation}
\bm{P^\prime}=\frac{\lbrack\bm{P} \cdot \bm{\hat n}+ S\!\left(\theta\right)\rbrack\bm{\hat n} + T\!\left(\theta\right)\bm{\hat n}\times\left(\bm{\hat n} \times \bm{P} \right)+U\!\left(\theta\right)\left(\bm{\hat n} \times \bm{P} \right)}{1+ S\!\left(\theta\right) \bm{P} \cdot \bm{\hat n}}
\label{pol1}
\end{equation}

with

\begin{equation}\nonumber
T\!\left(\theta\right)=\frac{{|f\!\left(\theta\right)|}^2-{|g\!\left(\theta\right)|}^2}{{|f\!\left(\theta\right)|}^2+{|g\!\left(\theta\right)|}^2}\,\,\,\,\text{and}\,\,\,\,U\!\left(\theta\right)=\frac{f\,g^{*}+f^{*}\,g}{{|f\!\left(\theta\right)|}^2+{|g\!\left(\theta\right)|}^2}\,.
\end{equation}

As for the treatment of elastic scattering in PEBSI, the ${\left( \frac{\text{d}\sigma}{\text{d}\Omega} \right)}_{0}$~values are taken from the NIST database of electron elastic-scattering cross sections (version 3.1)~\cite{nist1} while the $S$, $T$ and $U$ functions are derived from the Dirac equation using the Thomas-Fermi-Dirac atomic potential as described in~\cite{yas01}.

\subsection{Inelastic electron-electron scattering}
To model the inelastic scattering and also the energy loss of incident electrons we applied the commonly used approach of a hybrid model in which inelastic interactions are considered as to be either 'hard' or 'soft'. Hard interactions imply a significant amount of energy loss $\Delta E \ge E_\text{thresh}$ and angular deflection $\sin^2 \Delta \theta = \Delta E/E$, and are therefore explicitly treated as discrete collision events according to Eq.~\ref{eq1}. On the other hand soft collisions are considered as a continuous process with the energy loss being proportional to the distance traveled by the electron between two discrete interactions while the momentum and the spin orientation of the electrons remain unchanged. In this work we set $E_\text{thresh} = 1$\,keV. The stopping power due to soft interactions is then obtained from

\begin{equation}
\left(\frac{\text{d}E}{\text{d}s}\right)_{\text{soft}}=\,\left(\frac{\text{d}E}{\text{d}s}\right)_{\text{CSDA}}-\,\left(\frac{\text{d}E}{\text{d}s}\right)_{\text{hard}}\,,
\label{eloss1}
\end{equation}

where $\left(\frac{\text{d}E}{\text{d}s}\right)_{\text{CSDA}}$ is the CSDA (continuous slowing down approximation) stopping power provided by the ESTAR program~\cite{nist2} from NIST and $\left(\frac{\text{d}E}{\text{d}s}\right)_{\text{hard}}$ is calculated as

\begin{equation}
\left(\frac{\text{d}E}{\text{d}s}\right)_{\text{hard}}=\rho\!\int\limits_{\varepsilon_{\text{min}}}^{\varepsilon_{\text{max}}}E\varepsilon \left(\frac{\text{d}\sigma_{\text{inel}}}{\text{d}\varepsilon}\right)\text{d}\varepsilon
\label{eloss2}
\end{equation}

with the incident electron kinetic energy $E$ and the fractional energy loss $\varepsilon=\Delta E/E$. The minimum and the maximum energy loss for a specific interaction are given by $\varepsilon_{\text{min}}$ and $\varepsilon_{\text{max}}$, respectively. The lower threshold~$\varepsilon_{\text{min}}$ is defined as the maximum value of two parameters, namely the energy that distinguishes hard from soft collisions $E_\text{thresh}$ and the binding energy of a specific target electron, while the upper threshold~$\varepsilon_{\text{max}}$ equals either $0.5$ (free-free collisions) or $1$ (free-bound). The effective cross section $\sigma_{\text{inel}}$ is the sum of the individual inelastic interaction cross sections being multiplied by the number of respective target electrons as described in the following.

If the binding energy of the target electrons can be neglected the electron-electron scattering process is described by the M{\o}ller cross section for free electrons:

\begin{equation}
\frac{\text{d}\sigma_{\text{free}}}{\text{d}\varepsilon}=n_{\text{free}} C_1 \left(\frac{1}{\varepsilon^2}+\frac{1}{\left(1-\varepsilon\right)^2}-\frac{\left(\gamma-1\right)^2}{\gamma^2}+\frac{(2-4\gamma)}{2\gamma^2}\frac{1}{\varepsilon\left(1-\varepsilon\right)}\right)
\label{moller}
\end{equation}

with the prefactor $C_1$ defined as

\begin{equation}\nonumber
C_1=\frac{2\,\pi \left(r_{\text{e}}\gamma\right)^2}{\left(\gamma-1\right)^2 \left(\gamma+1\right)}\,.
\end{equation}

Here $r_{\text{e}}$ denotes the classical electrons radius, $\gamma$ is the relativistic factor of the incident electron and $n_{\text{free}}$ is the number of target electrons considered as quasi-free. This approximation was applied to all electrons with binding energies~$I_{i} < E_{\text{thresh}}$, resulting in $\varepsilon_{\text{min}} = E_{\text{thresh}}/E$. Note that besides the fact that Eq.~\ref{moller} is not applicable for collisions where the energy transfer is in the order of the binding energy, the lower threshold $\varepsilon_{\text{min}}>0$ in Eq.~\ref{eloss2} is also necessary because the M{\o}ller cross section leads to a divergence of the integral when $\varepsilon_{\text{min}} \rightarrow 0$. We have to stress that because of this limitations the PEBSI code is not able to reproduce subtle features that are connected to low-energetic interactions and can only be treated adequately when using much more sophisticated models, see, e.g., \cite{fer96} and~\cite{fer98}. However, as the energy resolution of the x-ray polarimeters mentioned above is in the order of 2\,keV such features will not significantly alter the properties of the detected bremsstrahlung.

\begin{figure}[h]
\begin{center}
\includegraphics[width=8.6cm]{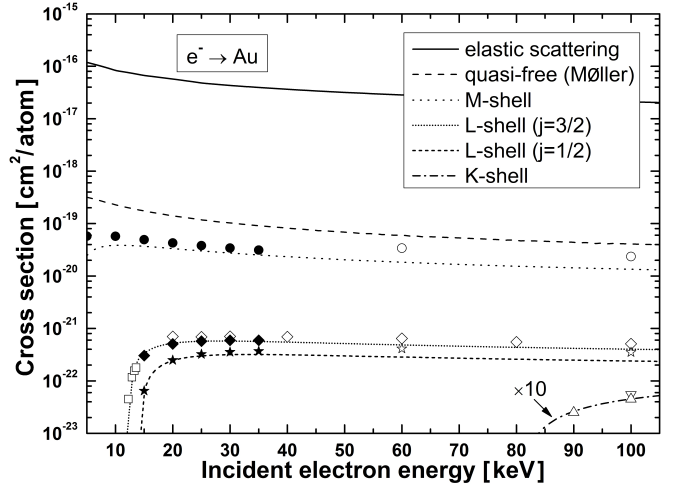}
\end{center}
\caption{Inelastic cross sections as a function of electron impact energy multiplied by the number of target electrons for collisions with gold atoms. The PEBSI cross sections are compared to various experimental and theoretical data: DWBA calculations -  $\medbullet$ \ding{117} \ding{72}~\cite{mer08}, experiment - $\medcircle$~\cite{pal80b} $\meddiamond$~\cite{dav72} $\medsquare$~\cite{shi81} $\medstar$~\cite{pal80a} $\medtriangleup$~\cite{dav72} $\medtriangledown$~\cite{mot64}.
\label{inel}}
\end{figure}

For target electron shells $i$ with binding energies~$I_{i} \ge E_{\text{thresh}}$ we apply the Gryzinski cross section~\cite{gry65}:

\begin{equation}
\begin{split}
\frac{\text{d}\sigma_{\text{Gry}}}{\text{d}\varepsilon}=n_i C_2 \frac{\varepsilon_i}{\varepsilon^3}{\left(1+\varepsilon_i\right)}^{-\frac{3}{2}}{\left(1-\varepsilon\right)}^{\varepsilon_i/\left(\varepsilon_i+\varepsilon\right)}\\\times\left(\frac{\varepsilon}{\varepsilon_i}\left(1-\varepsilon_i\right)+\frac{4}{3}\ln{\left(e+{\left(\frac{1-\varepsilon}{\varepsilon_i}\right)}^{\frac{1}{2}}\right)}\right)
\end{split}
\label{gryzinski}
\end{equation}

with $\varepsilon_i=I_i/E$ and the prefactor $C_2$ given by

\begin{equation}\nonumber
C_2=\frac{\pi\,{r_{\text{e}}}^2}{\left(\gamma-1\right)^2}\,R\,.
\end{equation}

The $R$ denotes a correction factor accounting for relativistic incident electron energies and/or relativistic binding energies which was also introduced by Gryzinski. Here, we use a slightly modified $R$ that was proposed in~\cite{haq10}. Though relying on a purely classical collision model, the electron impact ionization cross sections yielded by Eq.~\ref{gryzinski} are often in reasonable agreement with data from experiments and more sophisticated calculations. However, in the near-threshold region the Gryzinski model tends to significantly underestimate the total cross section values, see~\cite{nam99}. Therefore we replaced the total electron impact ionization cross section for the K-and the L-shell electrons by the semi-empirical formula given by Haque \etal{}~\cite{haq10,haq11}, while keeping the energy differential behavior given by Eq.~\ref{gryzinski}.

Fig.~\ref {inel} shows the resulting cross sections as a function of the incident electron energy. The PEBSI results are compared to data from DWBA calculations and several experiments. In general, reasonable agreement is found. Although the collisions with inner-shell electrons give a minor contribution to the electron transport as it is dominated by the elastic scattering process, the cross sections can be used to model the resulting fluorescence x-rays which may alter the bremsstrahlung radiation. In the future, we also might take into account the secondary electrons generated by electron-electron collisions in order to investigate their contribution to the bremsstrahlung.

Note that in case of electron-electron collisions, we do not consider the electron spin polarization as the scattered electron distribution is altered only in the case where both collision partners are polarized. However, the incident electron spin orientation is altered during the collision leading to a successive depolarization of the electron beam. At the moment this effect is taken into account by setting the electron to be unpolarized after the first hard collision. Naturally, this assumption overestimates the effect of depolarization and we plan to implement are more realistic treatment in the near future.

\subsection{Verification of the model}
In order test the reliability of our electron transport model we compare the PEBSI results with predictions by a different Monte Carlo code as well as with experimental data.

\begin{figure}[h]
\begin{center}
\includegraphics[width=8.6cm]{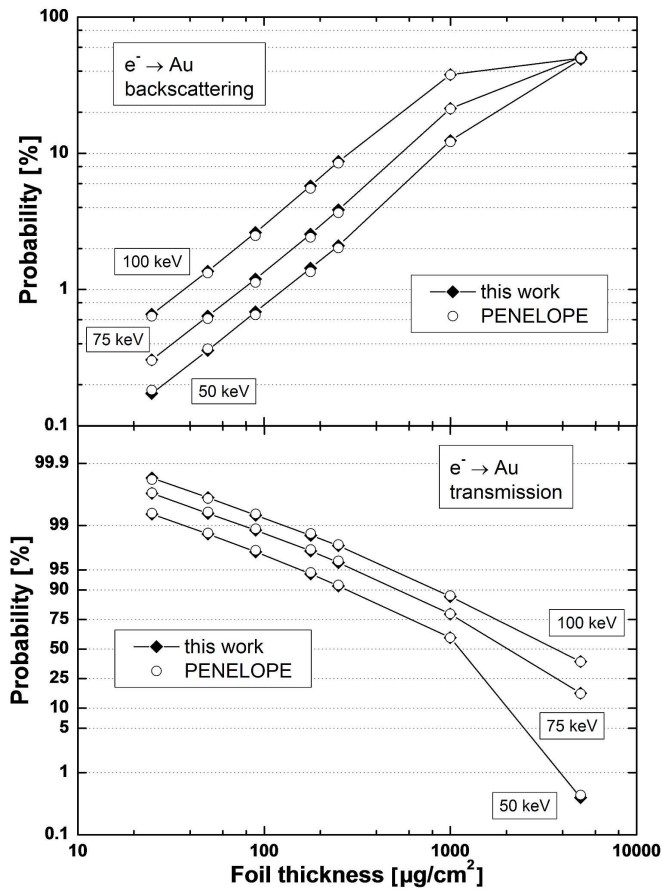}
\end{center}
\caption{Backscattering and transmission probability for electrons impinging on gold foils of various thicknesses. The PEBSI results are compared to predictions by the \mbox{PENELOPE} program~\cite{sal09}. The lines connecting the data points are drawn to guide the eye.
\label{pebsi_vs_pen}}
\end{figure}

In Fig.~\ref{pebsi_vs_pen} the percentage of backscattered and transmitted electrons is shown as a function of target thickness for different incident electron energies. We simulated an unpolarized electron beam impinging at normal incidence on gold foils. The data from PEBSI are compared to predictions by the \mbox{PENELOPE} code (version 2008)~\cite{sal09}, which is a general-purpose Monte Carlo program package for electron and photon transport and is known to provide a very reliable transport algorithm for unpolarized electrons~\cite{sem03}. In order to perform the comparison under almost identical conditions, both codes were adjusted to discard electrons with energies falling below $E_\text{cut}=1$\,keV and the minimum energy loss treated as a hard inelastic collision was set to 1\,keV~(PEBSI) or to 1\,\% of the electron energy (\mbox{PENELOPE}), respectively. Moreover, \mbox{PENELOPE} was set to treat the elastic scattering process as detailed as possible (by setting the maximum angular deflection that is treated with a multiple-scattering approach to zero). The generation of secondary electrons as well as the emission of bremsstrahlung was not taken into account. As seen, both programs yield nearly identical results.

\begin{figure}[h]
\begin{center}
\includegraphics[width=8.6cm]{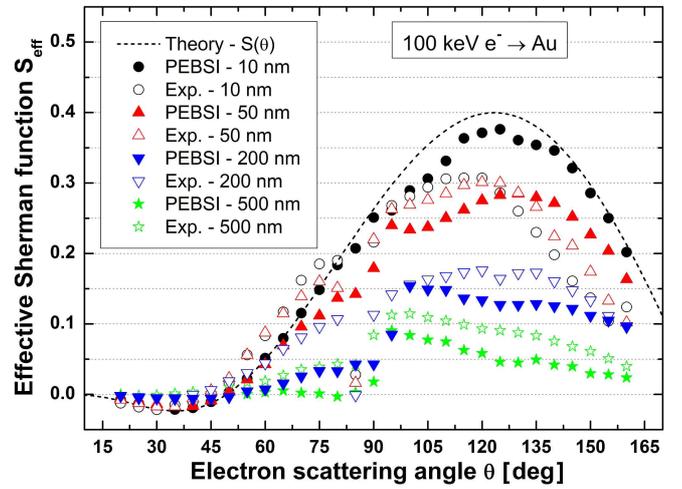}
\end{center}
\caption{Anisotropy parameter $S_\text{eff}$ of the scattered electron distribution in case of 100\,keV electrons being transversely polarized with respect to the electron momentum. The PEBSI results for gold targets of various thicknesses are compared to experimental data~\cite{koh06}.
\label{mott}}
\end{figure}

In case of electrons being transversely polarized with respect to the electron momentum, the azimuthal asymmetry of the elastically scattered electron distribution is characterized by the effective Sherman function $S_\text{eff}\!\left(\theta\right)$. For infinite-thin targets one expects that $S_\text{eff}$ equals $S$ and, consequently, the scatter distribution is given by Eq.~\ref{mott2} as discussed above. In Fig.~\ref{mott}, PEBSI results obtained for gold targets of various thicknesses are compared to experimental data which were measured for an electron energy of 100\,keV~\cite{koh06}. The theoretical Sherman function for single-collision conditions is shown in addition. Although one finds a qualitative agreement between the data from simulation and experiment, a quantitative analysis would require a much more refined simulation taking into account the complete experimental setup. For example, the decrease of the experimental $S_\text{eff}$ in case of the thinnest target and at backward angles might be due to electrons being backscattered from the beam dump behind the target foil and the dip near 90$^\circ$ was probably caused by the target holder. However, such a detailed study is beyond the scope of the present work.

\section{Simulation of bremsstrahlung linear polarization}
As discussed above, the emission of bremsstrahlung radiation is not implemented as a part of the electron transport code. Instead, we treat the electrons like they were permanently emitting bremsstrahlung photons while we ignore the effect on the electron properties. The procedure is as follows: At every discrete interaction point according to Eq.~\ref{eq1}, we calculate the observation angle of a virtual x-ray detector with respect to the actual electron momentum. In addition, we estimate the electron spin orientation with respect to the reaction plane. This information together with the electron energy is used to obtain the cross section and polarization properties of bremsstrahlung which was emitted along the straight path from the previous interaction point to the actual one and was heading in the direction of the x-ray detector. As the bremsstrahlung emission probability is proportional to the distance traveled, the path length between both interactions is used as weighting factor. The theoretical electron-nucleus bremsstrahlung data is taken from tables which were calculated within a fully relativistic treatment, see~\cite{yer10} for details. Note that we do not take into account the contribution of electron-electron bremsstrahlung, which is roughly a factor $Z$ smaller than the ordinary bremsstrahlung. Scattering and absorption processes of the bremsstrahlung photons inside the target are also neglected. The superposition of the radiation stemming from a large number of electrons along their complete tracks inside the target material then transforms the theoretical data for the single-collision case to effective bremsstrahlung properties taking into account the target effects as well as the position and solid-angle coverage of the virtual detectors.

\begin{figure}[h]
\begin{center}
\includegraphics[width=8.6cm]{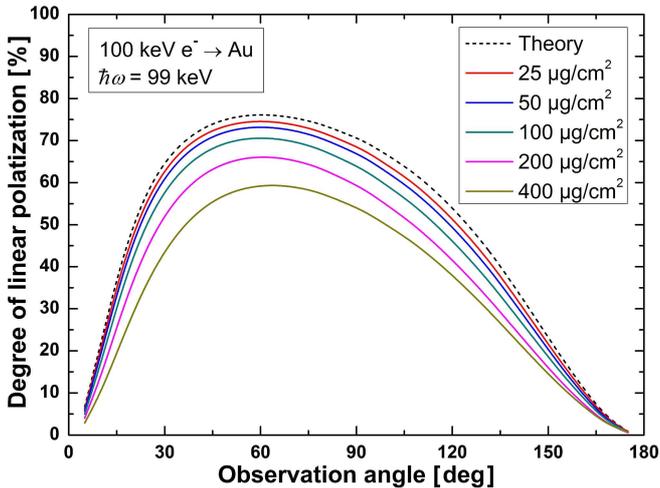}
\end{center}
\caption{The degree of linear polarization of the high energy end of bremsstrahlung plotted as a function of the observation angle and various target thicknesses. The PEBSI results for unpolarized electrons impinging on gold targets are compared to the underlying theoretical predictions for the single-collision case~\cite{yer10}.
\label{brems1}}
\end{figure}

As a first result, we present in Fig.~\ref{brems1} the degree linear polarization of bremsstrahlung stemming from unpolarized electrons impinging on gold foils for various target thicknesses in comparison to the underlying theoretical predictions for the single-collision case. The PEBSI results are shown for the high-energy end ($\hbar \omega / E=0.99$), also referred to as the short-wavelength limit, of the bremsstrahlung distribution where the polarization is most pronounced. In order to visualize target effects only, the virtual detectors as well as the electron beam radius were assumed to be point-like. A clear decrease in the degree of linear polarization is found even for the thinnest target foil considered. This finding is remarkable as for high-Z targets in the literature, thicknesses in the order of 50\,$\mu$g/cm$^2$ are regarded as thin enough in order to minimize the effect of multiple collisions on the emitted bremsstrahlung radiation~\cite{qua06}. As a consequence, one can in general expect significant target effects on the bremsstrahlung linear polarization for self-supporting \mbox{high-Z} target foils that are typically used in experiments.

\begin{figure}[h]
\begin{center}
\includegraphics[width=8.6cm]{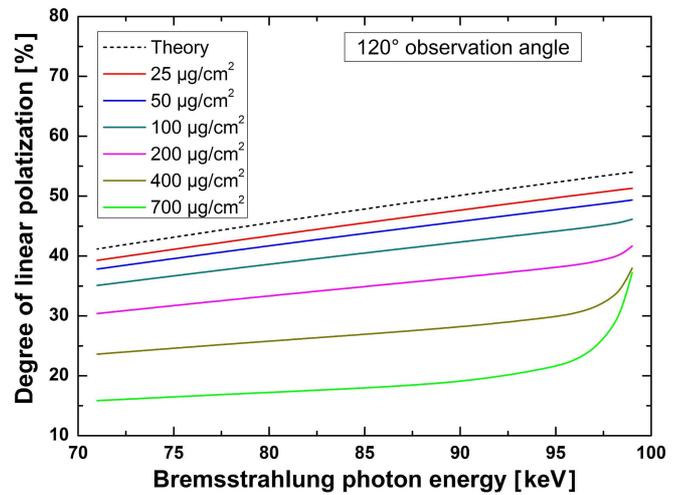}
\end{center}
\caption{Bremsstrahlung linear polarization as a function of the photon energy for an observation angle of 120$^\circ$ with respect to the direction of the incident electron beam, see text for details.
\label{brems2}}
\end{figure}

In fig.~\ref{brems2}, the degree of linear polarization is shown as a function of the bremsstrahlung photon energy for an observation angle of 120$^\circ$. For target thicknesses above 100\,$\mu$g/cm$^2$, the depolarization due to target effects decreases with increasing photon energy at the high-energy end of the bremsstrahlung distribution. This feature is caused by the fact that electrons with large deflection angles have such long tracks inside the target that their energy loss becomes significant and, consequently, after a certain track length these electrons have not enough kinetic energy to contribute to the highest bremsstrahlung energies. However, they can still emit bremsstrahlung with slightly lower photon energies and here their large deflection with respect to the incident electron direction leads to a strong depolarization when summing over the radiation emitted by a large number of individual electrons. This in contrast to thinner targets where only the small fraction of electrons with deflection angles very close to 90$^\circ$ (when assuming normal incidence of the incoming electrons) is likely to stay long enough inside the target to suffer a significant energy loss. From this results one can expect an almost complete depolarization of bremsstrahlung when high-Z targets with thicknesses in the order of 1\,$\mu$m~(equals 1932\,$\mu$g/cm$^2$ in the case of gold) are used and photon energies at least a few keV lower than the short-wavelength limit are observed.

As for the case of bremsstrahlung arising from polarized electrons, the development of the code is still ongoing. However, preliminary results already indicate that the PEBSI code is able to qualitatively reproduce the bremsstrahlung properties that were obtained in recent experiments.

\section{Acknowledgments}
We would like to thank R.~Barday, A.~K.~F.~Haque, D.~Jakubassa-Amundsen, T.~Kohashi and S.~Tashenov for the fruitful discussions during this work. A.~S. and V.~Y. also acknowledge the support by the Helmholtz Gemeinschaft (Nachwuchsgruppe VH-NG-421).

\label{}

\end{document}